\documentstyle [12pt,epsf] {article}

\catcode`@=12
\begin{document}
\thispagestyle{empty}
\begin{flushright} UCRHEP-T289\\October 2000\
\end{flushright}
\vspace{0.5in}
\begin{center}
{\Large LEPTOGENESIS: NEUTRINOS AND NEW LEPTON FLAVOR VIOLATION AT THE 
TEV ENERGY SCALE\\}
\vspace{1.2in}
{\bf Ernest Ma\\}
\vspace{0.2in}
{\sl Department of Physics, University of California, 
Riverside, CA 92521, USA} 
\vspace{1.2in}
\end{center}
\begin{abstract}\
Leptogenesis, i.e. the creation of a lepton asymmetry 
in the early Universe, may occur through the decay of heavy singlet 
(right-handed) neutrinos.  If we require it not to be erased by physics 
beyond the Standard Model at the TeV energy scale, then only 2 candidates 
are possible if they are subgroups of $E_6$.  These 2 solutions happen to 
be also the only ones within 1$\sigma$ of the atomic parity violation data 
and the invisible $Z$ width.  Lepton flavor violation is predicted in one 
model, as well as in another unrelated model of neutrino masses where 
the observable decay of a doubly charged scalar would determine the relative 
magnitude of each element of the neutrino mass matrix.
\end{abstract}
\vspace{0.1in}
--------------

\noindent Talk given at the Joint U.S./Japan Workshop on New Initiatives in 
Lepton Flavor Violation and Neutrino Oscillations with Very Intense Muon and 
Neutrino Sources (Honolulu, HI), October 2-6, 2000.

\newpage
\baselineskip 24pt

\section{Introduction}

In the minimal Standard Model, leptons transform under $SU(3)_C \times 
SU(2)_L \times U(1)_Y$ according to
\begin{equation}
\left( \begin{array} {c} \nu \\ l \end{array} \right)_L \sim (1,2,
-{1 \over 2}), ~~~ l_R \sim (1,1,-1).
\end{equation}
In the absence of $\nu_R$, the Majorana mass of the neutrino must come 
from the effective dimension-5 operator \cite{wema}
\begin{equation}
{1 \over \Lambda} (\nu_i \phi^0 - l_i \phi^+)(\nu_j \phi^0 - l_j \phi^+).
\end{equation}
This means that the so-called ``seesaw'' structure, i.e. $m_\nu \sim v^2/
\Lambda$ is inevitable, no matter what specific mechanism is used to obtain 
$m_\nu$.

The canonical seesaw mechanism \cite{seesaw} is achieved with the addition 
of a heavy $N_R \sim (1,1,0)$.  In that case, the interaction $f \bar N_R 
\nu_i \phi^0$ and the large Majorana mass $m_N$ of $N_R$ allow the above 
effective operator to be realized, with
\begin{equation}
m_\nu = {f^2 \langle \phi^0 \rangle^2 \over m_N} = {m_D^2 \over m_N}.
\end{equation}

\section{Leptogenesis from $N$ Decay}

Consider the decay of $N$ in the early Universe. \cite{fuya,flpasa} Since it 
is a heavy Majorana particle, it can decay into both $l^- \phi^+$ (with 
lepton number $L=1$) and $l^+ \phi^-$ (with $L=-1$).  Hence $L$ is violated.  
Now $CP$ may also be violated if the one-loop corections are taken into 
account.  Specifically, consider $N_1 \to l^- \phi^+$.  This amplitude has 
contributions from the tree diagram as well as a vertex correction and a 
self-energy correction, with $l^+ \phi^-$ in the intermediate state and 
$N_{2,3}$ appearing in the cross and direct channels respectively.  Calling 
this amplitude $A + iB$, where $A$ and $B$ are the dispersive and absorptive 
parts, the asymmetry generated by $N_1$ decay is then proportional to
\begin{equation}
|A + iB|^2 - |A^* + iB^*|^2 = 4 {\rm Im} (A^*B),
\end{equation}
which is nonzero if $A$ and $B$ have a relative phase, i.e. if $CP$ is 
violated.  Note that if there is only one $N$ (i.e. $N_{2,3}$ exchange is 
absent), then this phase is automatically zero in the above.

In the approximation that $M_1$ is much smaller than $M_{2,3}$, the decay 
asymmetry is
\begin{equation}
\delta \simeq {G_F \over 2 \pi \sqrt 2} {1 \over (m_D^\dagger m_D)_{11}} 
\sum_{j=2,3} {\rm Im} (m_D^\dagger m_D)^2_{1j} {M_1 \over M_j},
\end{equation}
which may be washed out by the inverse interactions which also violate $L$ 
unless the decay occurs out of equilibrium with the rest of the particles 
in the Universe as it expands.  This places a constraint on $M_1$ to be 
in the range $10^9$ to $10^{13}$ GeV.

Once the $N$'s have decoupled as the Universe cools, the other (light) 
particles, i.e. those of the Standard Model, have only $L$ conserving 
interactions except for the nonperturbative sphalerons which violate 
$B + L$, but conserves $B - L$.  Hence the $L$ asymmetry generated by 
$N$ decay gets converted \cite{kurush} into a baryon asymmetry of the 
Universe, which is observed at present to be of order $10^{-10}$.

If $N$ decay is indeed the source of this $B$ asymmetry (to which we owe 
our own very existence), then any TeV extension of the Standard Model 
should also conserve $B - L$.  In the next section it will be shown that 
if this extension involves a subgroup of $E_6$, then there are only 2 
possible candidates. \cite{hmrs}

\section{Possible $E_6$ Subgroups at the TeV Scale}

Consider the maximal subgroup $SU(3)_C \times SU(3)_L \times SU(3)_R$ of 
$E_6$.  The fundamental \underline {27} representation is given by
\begin{equation}
\underline {27} = (3,3,1) + (3^*,1,3^*) + (1, 3^*,3).
\end{equation}
The fermions involved are all taken to be left-handed and defined to be
\begin{eqnarray}
&& (u,d) \sim (3;2,1/6;1,0), ~~~ h \sim (3;1,-1/3;1,0), \\ 
&& (d^c, u^c) \sim (3^*;1,0;2,-1/6), ~~~ h^c \sim (3^*;1,0;1,1/3), \\ 
&& (\nu_e, e) \sim (1;2,-1/6;1,-1/3), ~~~ (e^c, N) \sim (1;1,1/3;2,1/6), \\ 
&& (E^c, N_E^c), (\nu_E, E) \sim (1;2,-1/6;2,1/6), ~~~ S \sim (1;1,1/3;1,
-1/3),
\end{eqnarray}
under $SU(3)_C \times SU(2)_L \times U(1)_{Y_L} \times SU(2)_R \times 
U(1)_{Y_R}$.  In this notation, the electric charge is given by $Q = T_{3L} 
+ Y_L + T_{3R} + Y_R$, with $B-L = 2(Y_L + Y_R)$.

Since $(e^c, N)$ is an $SU(2)_R$ doublet, the requirement that $m_N > 10^9$ 
GeV for successful leptogenesis is not compatible with the existence of 
$SU(2)_R$ at the TeV scale.  This rules out the subgroup $SU(3)_C \times 
SU(2)_L \times SU(2)_R \times U(1)_{B-L}$ of $SO(10)$.  However, as shown 
below, a different decomposition of $SU(3)_R$, i.e. into $SU(2)'_R \times 
U(1)_{Y'_R}$, with
\begin{equation}
T'_{3R} = {1 \over 2} T_{3R} + {3 \over 2} Y_R,  ~~~ Y'_R = {1 \over 2} 
T_{3R} - {1 \over 2} Y_R,
\end{equation}
allows $N$ to be trivial under the new skew left-right gauge group \cite{ma87} 
so that its existence at the TeV scale is compatible with $N$ leptogenesis.

To see how this works, consider the decomposition of $E_6$ into its 
$SO(10)$ and $SU(5)$ subgroups, then
\begin{eqnarray}
\underline {27} &=& (16,5^*)[d^c,\nu_e,e] + (16,10)[u,d,u^c,e^c] + (16,1)[N] 
\nonumber \\ &+& (10,5^*)[h^c,\nu_E,E] + (10,5)[h,E^c,N_E^c] + (1,1)[S].
\end{eqnarray}
If we now switch $(16,5^*)$ with $(10,5^*)$ and (16,1) with (1,1), then the 
$SU(5)$ content remains the same, but the $SO(10)$ does not.  The result 
is a different choice of the direction of $SU(3)_R$ breaking, i.e. $V$ spin 
instead of the usual $T$ spin.  Specifically, we switch $d^c$ with $h^c$, 
$(\nu_e, e)$ with $(\nu_E, E)$, and $N$ with $S$ in Eqs.(8) to (10).  Now 
we may let $N$ be heavy without affecting the new skew left-right gauge 
group
\begin{equation}
SU(3)_C \times SU(2)_L \times SU(2)'_R \times U(1)_{Y_L + Y'_R}.
\end{equation}
Note that $B-L$ is conserved by all the interactions of this model at 
the TeV scale.

Consider next the decomposition $E_6 \to SO(10) \times U(1)_\psi$, then 
$SO(10) \to SU(5) \times U(1)_\chi$, where
\begin{equation}
Q_\psi = \sqrt {3 \over 2} (Y_L - Y_R), ~~~ Q_\chi = \sqrt {1 \over 10} 
(5 T_{3R} - 3Y).
\end{equation}
The arbitrary linear combination $Q_\alpha \equiv Q_\psi \cos \alpha + Q_\chi 
\sin \alpha$ has been studied extensively as a function of $\alpha$.  If we 
let $\tan \alpha = 1/\sqrt {15}$, then \cite{ma96}
\begin{equation}
Q_N = \sqrt {1 \over 40} (6 Y_L + T_{3R} - 9 Y_R).
\end{equation}
In that case, $N$ is also trivial under this $U(1)_N$.  Hence
\begin{equation}
SU(3)_C \times SU(2)_L \times U(1)_Y \times U(1)_N
\end{equation}
is the second and only other possible $E_6$ extension of the Standard Model 
compatible with $N$ leptogenesis.

\section{New Neutral Currents and Lepton Flavor Violation}

In the $SU(3)_C \times SU(2)_L \times SU(2)'_R \times U(1)_{Y_L + Y'_R}$ 
model, $(h^c, u^c)$ and $(e^c, S)$ are $SU(2)'_R$ doublets, but whereas 
$u^c$ has $B-L = -1/3$, $h^c$ has $B-L = 2/3$, and whereas $e^c$ has 
$B-L = 1$, $S$ has $B-L = 0$, hence the $W_R^-$ gauge boson of this model 
has $B-L =-1$ (because $T'_{3R} = -1$ and $Y'_R = 0$ imply $Y_R = -1/2$). 
This unusual property has been studied extensively.  Moreover, if $S$ is 
light, it may be considered a ``sterile'' neutrino.  In that case, it 
has recently been shown \cite{ma00} that $M_{W_R} > 442$ GeV.

The extra neutral gauge boson $Z'$ of this model is related to $W_R$ by
\begin{equation} 
M_{Z'} = (\cos \theta_W / \cos 2 \theta_W) M_{W_R} > 528 ~{\rm GeV},
\end{equation}
and it couples to \cite{bahema}
\begin{eqnarray}
&& {1 \over \sqrt {1-2x}} [xT_{3R} + (1-x)T'_{3R} - xQ] \nonumber \\ &=& 
{-1 \over \sqrt {1-2x}} [xY_L + \left( {3x-1 \over 2} \right) T_{3R} - 
\left( {3-5x \over 2} \right) Y_R],
\end{eqnarray}
where $x \equiv \sin^2 \theta_W$ and $g_L = g_R$.  The $Z$ boson of this 
model behaves in the same way as that of the Standard Model, except
\begin{equation}
Z = \sqrt {1-x} W_L^0 - {x \over \sqrt {1-x}} W_R^0 - {\sqrt x \sqrt {1-2x} 
\over \sqrt {1-x}} B,
\end{equation}
which implies a $Z W_R^+ W_R^-$ coupling that is absent in the Standard 
Model.

Together with the $Z'$ of the $U(1)_N$ model, the extra neutral-current 
interactions of these two $E_6$ subgroups are the only ones within 1$\sigma$ 
of the atomic parity violation data \cite{bewi} and the invisible $Z$ 
width. \cite{erla}  [The $U(1)_N$ model was not considered in Ref.~[12], 
but it can easily be included in their Fig.~1 by noting that it has 
$\alpha = 0$ and $\tan \beta = \sqrt {15}$ in their notation.]  The 
remarkable convergence of the requirement of successful $N$ leptogenesis 
and the hint from present neutral-current data regarding possible new 
physics at the TeV scale is an encouraging sign for the validity of one 
of these models.

Because of the $ZW_R^+W_R^-$ coupling, lepton flavor violation occurs in 
one loop through the effective $Z \bar e \mu$ vertex.  This is the analog 
of the $ZW_L^+W_L^-$ contribution in the Standard Model.  The latter is 
negligible because all the neutrino masses are very small; the former is 
not because $m_{S_3} = M_{Z'}$ in the simplest supersymmetric version of this 
model. \cite{ma87,ma00,bahema}  The effective $\mu-e$ transition coupling is 
then given by
\begin{equation}
g_{Z \bar e \mu} = {e^3 U_{\mu3} U_{e3} \over 16 \pi^2 \sqrt {x(1-x)}} 
\left[ {r_3 \over 1-r_3} + {r_3^2 \ln r_3 \over (1-r_3)^2} \right],
\end{equation}
where $r_3 = m_{S_3}^2/M_{W_R}^2 = 1.426$ and $S_{1,2}$ are assumed light.

Using present experimental bounds, upper limits of the mixing of $S_3$ to 
$\mu$ and $e$ are given below.
\begin{eqnarray}
&& |U_{\mu3}U_{e3}| < 2.3 \times 10^{-3} ~{\rm from}~ B(\mu\to eee) < 1.0 
\times 10^{-12}; \\ 
&& |U_{\mu3}U_{e3}| < 3.6 \pm 0.9 \times 10^{-3} 
~[\mu-e ~{\rm conversion~in}~ ^{48}Ti,~ ^{208}Pb]; \\ 
&& |U_{\mu3}U_{e3}| {M_{W_L}^2 \over M_{W_R}^2} < 3.8 \times 10^{-4} 
~{\rm from}~ B(\mu \to e \gamma) < 1.2 \times 10^{-11}.
\end{eqnarray}
This shows that unless the mixing angles are extremely small, future 
precision experiments on lepton flavor violation will be able to test 
this model in conjunction with TeV colliders.

\section{New Verifiable Model of Neutrino Masses}

Let us go back to the effective operator of Eq.~(2) and rewrite it as
\begin{equation}
{1 \over \Lambda} [\nu_i \nu_j \phi^0 \phi^0 - (l_i \nu_j + \nu_i l_j) \phi^0 
\phi^+ + l_i l_j \phi^+ \phi^+].
\end{equation}
This tells us that another natural realization of a small Majorana neutrino 
mass is to insert a heavy scalar triplet $\xi = (\xi^{++},\xi^+,\xi^0)$  
with couplings to leptons
\begin{equation}
f_{ij} [\xi^0 \nu_i \nu_j + \xi^+ (\nu_i l_j + l_i \nu_j)/\sqrt 2 + \xi^{++} 
l_i l_j] + h.c.,
\end{equation}
and to the standard scalar doublet
\begin{equation}
\mu [\bar \xi^0 \phi^0 \phi^0 - \sqrt 2 \xi^- \phi^+ \phi^0 + \xi^{--} \phi^+ 
\phi^+] + h.c.
\end{equation}
We then obtain \cite{masa}
\begin{equation}
m_\nu = {2 f_{ij} \mu \langle \phi^0 \rangle^2 \over m_\xi^2} = 2 f_{ij} 
\langle \xi^0 \rangle.
\end{equation}
This shows the inevitable seesaw structure, but instead of identifying 
$m_N$ with the large scale $\Lambda$ as in the canonical seesaw model 
\cite{seesaw}, we now require only $m_\xi^2/\mu$ to be large.  If $\mu$ is 
sufficiently small, the intriguing possibility exists for $m_\xi$ to be 
of order 1 TeV and be observable at future colliders.  The decay
\begin{equation}
\xi^{++} \to l_i^+ l_j^+
\end{equation}
is easily detected and its branching fractions {\bf determine} the relative 
$|f_{ij}|$'s, i.e. the $3 \times 3$ neutrino mass matrix up to phases 
and an overall scale. \cite{marasa}  This possible connection between 
collider phenomenology and neutrino oscillations is an extremely attractive 
feature of the proposed Higgs triplet model of neutrino masses.  

To understand why $\mu$ can be so small and why $m_\xi$ should be of order 
1 TeV, one possibility \cite{marasa} is to consider the Higgs triplet model 
in the context of large extra dimensions.  $\mu$ is small here because it 
violates lepton number and may be represented by the ``shining'' of a singlet 
scalar in the bulk, i.e. its vacuum expectation value as felt in our brane. 
$m_\xi$ is of order 1 TeV because it should be less than the fundamental 
scale $M_*$ in such theories which is postulated to be of order a few TeV.

Lepton flavor violation in this model may now be predicted if we know 
$f_{ij}$.  Using a hierarchical neutrino mass matrix which fits present 
atmospheric \cite{atm} and solar \cite{sol} neutrino oscillations (choosing 
the large-angle MSW solution), we predict \cite{marasa} $\mu-e$ conversion 
to be easily observable at the MECO experiment as shown in 
Fig.~1 if $m_\xi$ is indeed of order 1 TeV.  The dimensionless parameter 
$h$ there is proportional to $\mu$.

\begin{figure}[t]
\begin{center}
\epsfxsize = 1.0 \textwidth \epsffile{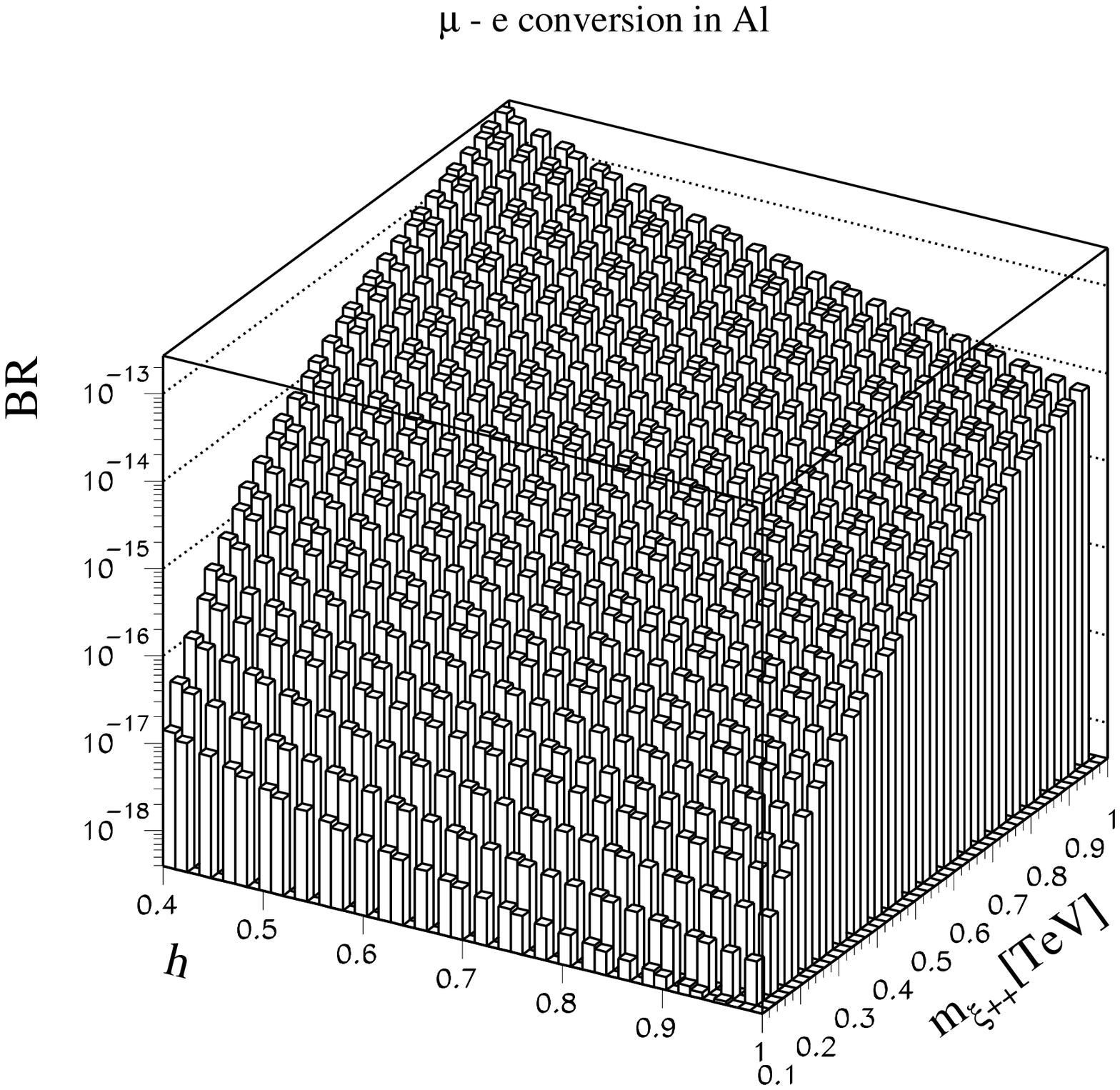}
\caption{Rate of $\mu-e$ conversion in $^{13}Al$.}
\end{center}
\end{figure}

\section{Conclusion}

Leptogenesis, neutrino masses, lepton flavor violation, and new physics at 
future colliders are most likely intertwined.  They may well be the different 
colors of a rainbow ({\bf manoa}) and must exist {\bf together} or not at all.

\section*{Acknowledgments}

I thank Y. Kuno and W. Molzon for a very useful and stimulating Workshop.
This work was supported in part by the U.~S.~Department of Energy under 
Grant No.~DE-FG03-94ER40837.


\begin{thebibliography}{99}

\bibitem{wema} S. Weinberg, Phys. Rev. Lett. {\bf 43}, 1566 (1979); E. Ma, 
Phys. Rev. Lett. {\bf 81}, 1171 (1998).

\bibitem{seesaw} T. Yanagida, in {\em Proceedings of the Workshop on the 
Unified Theory and the Baryon Number in the Universe}, edited by O. Sawada 
and A. Sugamoto (KEK Report No.~79-18, Tsukuba, Japan, 1979), p.~95; 
M. Gell-Mann, P. Ramond, and R. Slansky, in {\em Supergravity}, edited by 
P. van Nieuwenhuizen and D. Z. Freedman (North-Holland, Amsterdam, 1979), 
p.~315; R. N. Mohapatra and G. Senjanovic, Phys. Rev. Lett. {\bf 44}, 1316 
(1980).

\bibitem{fuya} M. Fukugita and T. Yanagida, Phys. Lett. {\bf 174B}, 45 (1986).

\bibitem{flpasa} M. Flanz, E. A. Paschos, and U. Sarkar, Phys. Lett. 
{\bf B345}, 248 (1995); Erratum: {\bf B382}, 447 (1996).

\bibitem{kurush} V. A. Kuzmin, V. A. Rubakov, and M. E. Shaposhnikov, Phys. 
Lett. {\bf 155B}, 36 (1985).

\bibitem{hmrs} T. Hambye, E. Ma, M. Raidal, and U. Sarkar, in preparation.

\bibitem{ma87} E. Ma, Phys. Rev. {\bf D36}, 274 (1987).

\bibitem{ma96} E. Ma, Phys. Lett. {\bf B380}, 286 (1996).

\bibitem{ma00} E. Ma, hep-ph/0002284 (Phys. Rev. {\bf D}, in press).

\bibitem{bahema} K. S. Babu, X.-G. He, and E. Ma, Phys. Rev. {\bf D36}, 878 
(1987).

\bibitem{bewi} S. C. Bennett and C. E. Wieman, Phys. Rev. Lett. {\bf 82}, 
2484 (1999).

\bibitem{erla} J. Erler and P. Langacker, Phys. Rev. Lett. {\bf 84}, 212 
(2000).

\bibitem{masa} E. Ma and U. Sarkar, Phys. Rev. Lett. {\bf 80}, 5716 (1998).

\bibitem{marasa} E. Ma, M. Raidal, and U. Sarkar, hep-ph/0006046 (Phys. 
Rev. Lett., in press). 

\bibitem{atm} H. Sobel, Talk given at {\it Neutrino 2000}, Sudbury, Canada 
(June 2000).

\bibitem{sol} Y. Suzuki, Talk given at {\it Neutrino 2000}, Sudbury, Canada 
(June 2000).

\end{thebibliography}
\end{document}